\documentclass[12pt]{article}
\usepackage{amssymb,amsmath}
\usepackage{graphicx}
\usepackage{color}
\newcommand{\diag}{\mathrm{diag}}
\newcommand{\Ncal}{\mathcal{N}}

\newcommand{\nbox}{{\,\lower0.9pt\vbox{\hrule \hbox{\vrule height 0.2 cm \hskip 0.19 cm \vrule height 0.2 cm}\hrule}\,}}

\def\href#1#2{#2}

\begin{document}
\begin{titlepage}
\hfill
\vbox{
    \halign{#\hfil         \cr
           hep-th/0701277  \cr
           } % end of \halign
      }  % end of \vbox
\vspace*{20mm}
\begin{center}
{\Large \bf General Lin-Maldacena solutions and\\
 PWMM Instantons from supergravity}

\vspace*{15mm}
\vspace*{1mm}

{Greg van Anders}

\vspace*{1cm}

{Department of Physics and Astronomy,
University of British Columbia\\
6224 Agricultural Road,
Vancouver, B.C., V6T 1Z1, Canada}

\vspace*{1cm}
%%\maketitle
\end{center}

\begin{abstract}
We use the Lin-Maldacena prescription to demonstrate how to find the
supergravity solutions dual to arbitrary vacua of the plane wave matrix model
and maximally supersymmetric Yang-Mills theory on $R \times S^2$, by solving
the auxiliary electrostatics problem. We then apply the technique to study
instantons at strong coupling in the matrix model.
\end{abstract}

\end{titlepage}

\vskip 1cm
\section{Introduction}
The AdS/CFT correspondence \cite{adscft} has proven to be a remarkably useful
tool for studying $\Ncal=4$ SYM at strong coupling. However, this is only
the most well-known example of gauge/gravity duality. Lin and Maldacena
\cite{lm}, building on the work of \cite{llm}, have found another very
interesting class of dualities. This class relates $SU(2|4)$ supersymmetric
gauge theories (the plane wave matrix model, and maximally supersymmetric
Yang-Mills theories on $R\times S^2$ and $R\times S^3/Z_k$) to type IIA string
theory in various backgrounds. The metric, dilaton, and form fields in
the supergravity solutions in question can be expressed in terms of a single
function that is an axisymmetric solution of the Laplace equation in three
dimensions. The gravity solutions, therefore, are specified by axisymmetric
electrostatics configurations. The field theories in question have many vacua,
and Lin and Maldacena \cite{lm} were able to determine the correspondence
between the vacua and the supergravity solutions. These present interesting
examples of the gauge/gravity correspondence as field theories living in
different numbers of dimensions are dual to string theory in backgrounds that
share similar features in the infrared region, such as throats with NS5- or
D2-brane flux.

The electrostatics configurations corresponding to the field theory vacua
are given by different arrangements of charged conducting disks \cite{lm}.
For a general vacuum on the field theory side, and therefore a general disk
configuration on the gravity side, solving the electrostatics problem is quite
challenging, and explicit solutions are known in only some special cases.

One particular solution given by Lin and Maldacena \cite{lm} corresponds to two
infinitely large disks held at fixed separation. They gave an explicit form for
the gravity solution and argued that it should be dual to little string theory
on $S^5$. The gravity dual was used in \cite{eb} to argue that little string
theory on $S^5$ has interesting features that differ from the theory in flat
space. An explicit solution has also been given in the case of a single
isolated disk, dual to a vacuum of the maximally supersymmetric Yang-Mills
theory on $R\times S^2$ \cite{lm,lmsvv}. Also, in the region very close to the
tip of a disk, the problem becomes two dimensional, and it is possible to solve
it by conformal mapping \cite{lm}. More general explicit solutions, however, are
not known.

In general this has prevented this set of dualities from being used to study
the $SU(2|4)$ symmetric field theories at strong coupling. It is interesting to
consider what questions can be addressed from the information we do know on
the gravity side. Recently, Lin \cite{lin} has made some progress in applying
this correspondence to instanton calculations in the plane wave matrix model
and maximally supersymmetric Yang-Mills theory on $R\times S^2$. In the case
of the matrix model, this question had been studied directly in the field
theory \cite{yy}. Lin gave explicit results for weak coupling from the
supergravity side, and found precise agreement with the gauge theory analysis.

It would certainly be desirable to be able to perform other gauge theory
calculations using the dual gravity description. It is, therefore, quite
interesting to obtain more general supergravity solutions that would allow this
to be done.

In this paper we will demonstrate that it is possible to reduce the generic
electrostatics problem to a simple linear system that can be solved very
simply using numerical methods. We will then use this technique to find some
explicit results using Lin's prescription for instanton calculations on the
dual gravity side. For a simple example electrostatics configuration, dual to
a vacuum of the plane wave matrix model, we will give an explicit expression
for the superpotential at strong coupling, and also the leading correction to
Lin's result at weak coupling.

\section{Supergravity solutions} \label{sugrasolns}
In this section we will review the Lin-Maldacena formulation of supergravity
solutions in terms of electrostatics problems, full details can be found in
\cite{lm}. Then in subsections \ref{gen} and \ref{gpwmm} we will discuss the
solution of these problems.

To find the supergravity duals to field theories with $SU(2|4)$ symmetry, Lin
and Maldacena looked for similarly symmetric supergravity solutions. In
particular, the bosonic part of this symmetry group is
$R\times SO(3)\times SO(6)$ so the supergravity solutions should contain an
$S^2$ and an $S^5$. Interestingly, with this restriction all of the
supergravity fields can be expressed in terms of a single function of the two
remaining coordinates. For the supergravity equations to be satisfied, this
function must be an axisymmetric solution to the Laplace equation in three
dimensions.

\begin{figure}
\begin{center}
\includegraphics{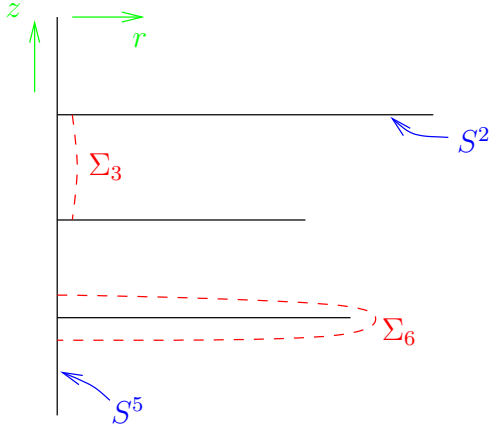}
\end{center}
\caption{An example electrostatics configuration. The conducting disks are
the horizontal solid lines. Fibred above the $rz$-plane are an $S^2$ and
an $S^5$. The size of the $S^2$ shrinks on the disks, whereas the $S^5$
shrinks on the $z$ axis. The dashed lines indicate topological 3- and 6-cycles
in the geometry.}
\label{cycles}
\end{figure}

The full supergravity solution in terms of this function, in the string frame,
is
\begin{eqnarray}
ds^2 &=& \left( \frac{\ddot V-2\dot V}{-V''} \right)^{1/2}
\left( - \frac{4\ddot V}{\ddot V-2\dot V} dt^2 + \frac{-2V''}{\dot V}
	(dr^2 + dz^2 )+4d\Omega_5^2 + 2\frac{V'' \dot V}{\Delta}
	d\Omega_2^2 \right), \cr
e^{4\Phi} &=& \frac{4(\ddot V-2\dot V)^3}{-V'' \dot V^2 \Delta^2}, \cr
C_1 &=&-\frac{2\dot V' \, \dot V}{\ddot V-2\dot V}dt,
	\label{sugrasoln} \\
F_4 &=&d C_3 ,\quad \quad \quad \quad \quad
	C_3=-4\frac{\dot V^2 V''}{\Delta} dt\wedge d^{2}\Omega , \cr
H_3 &=&d B_2 , \quad \quad \quad \quad \quad
	B_2=2 \left( \frac{\dot V\dot V'}{\Delta}+z \right) d^{2}\Omega, \cr
\Delta &\equiv& (\ddot V-2\dot V)V''-(\dot V')^2,
\nonumber
\end{eqnarray}
where $V$ is the electrostatics potential, and dots and primes indicate
derivatives with respect to $\log r$ and $z$, respectively. To avoid conical
singularities in \eqref{sugrasoln}, when the size of the $S^2$ or $S^5$ shrinks,
requires that either $V$ is regular at $r=0$ ($S^5$ shrinks) or $\partial_rV=0$
($S^2$ shrinks). Different supergravity solutions can, therefore, be specified
by inserting some conducting disks of various radii $R_i$ at positions $z_i$.
See figure \ref{cycles}. Inserting a disk will create separate two regions
on the $z$-axis on which the $S^5$ shrinks and will therefore mean adding
a non contractible 6-cycle, which will carry NS5-brane flux. Similarly, the
region between two disks, on which the $S^2$ shrinks, will be a non-contractible
3-cycle carrying D2-brane flux.

Two additional constraints on the electrostatics solution come from ensuring
that all of the metric components are positive definite and that the transformation to these coordinates is well defined. Positive definiteness requires that
the electrostatics potential takes a definite asymptotic form, and the
coordinate transformation requires that the charge density vanishes at the edge
of each disk.

We will now describe a method for solving the electrostatics problems for
generic configurations.

\subsection{General solutions dual to SYM on $R\times S^2$} \label{gen}
One of the field theories for which Lin and Maldacena found the corresponding
electrostatics configurations is maximally supersymmetric $SU(N)$ Yang-Mills
theory on $R\times S^2$. This theory is related to $\Ncal=4$ SYM on
$R\times S^3$ by dimensionally reducing that theory on the Hopf fibre of $S^3$
\cite{lm}. The field content is similar to $\Ncal=4$, however the theory on
$R\times S^2$ admits vacuum configurations with non-trivial $\Phi$, the scalar
field resulting from the dimensional reduction. The vacua of this theory are
parametrized by a set of integers where $\Phi=\diag(n_1, n_2, \ldots, n_N)$ 
\cite{msv}. General discussion of the relations among the vacua in the
$SU(2|4)$ symmetric field theories can be found in \cite{itt,istt}.

The set of electrostatics configurations in question are given by an arbitrary
set of positively charged conducting disks. Each disk will be associated with
some D2 brane charge, so it is natural to think of the supergravity solutions
as being dual to vacua of maximally supersymmetric Yang-Mills on $R\times S^2$
\cite{lm}. The integers in the vacuum configuration for $\Phi$ are related to
positions of the disks by $z = n\pi/2$, and the number of units of charge on the
disk is related to the number of times that each integer appears by
$Q=\pi^2 N/8$ \cite{lm}.

The solutions to these electrostatics problems have been given in some specific
cases \cite{lm,lmsvv,lsv}. For example the limit that the disks are very large,
or only the geometry near the tip of a disk is of interest, the problem becomes
two dimensional and it is possible to treat it with conformal
mapping \cite{lm,lmsvv,lin,lsv}. In the case that there is a single disk it is
possible to find an exact solution \cite{lm,lmsvv}. For two equally sized
disks a formal solution can be found \cite{lmsvv,lsv}. We will show that
the techniques of \cite{lmsvv,lsv,sneddon} to solve the electrostatics problem
for two identical disks can be extended to more arbitrary disk configurations.
We will discuss how these more general solutions may be found using these
techniques.

Consider the case of a collection of $k$ charged conducting disks in the case of
maximally supersymmetric Yang-Mills theory on $R\times S^2$. This problem is
similar to the one for two disks considered in \cite{lsv}, however we will allow
the disks here to sit at arbitrary positions, $d_i$ and have arbitrary sizes,
$R_i$. We can take the potential to be
\begin{equation}
V = W_0 \left( r^2-2z^2 + \sum_i \phi_i(r,z) \right),
\end{equation}
where the first two terms ensure the correct asymptotic conditions, and the
third is an asymptotically vanishing contribution that comes from the charges
on the disks. It takes the form
\begin{equation} \label{phid}
\phi_i(r,z) = \int_0^\infty \frac{du}{u} J_0(ru) A_i(u) e^{-u|z-d_i|} \, .
\end{equation}
Each function $A_i$ will be shown to determine the charge density on the
$i^{th}$
disk. To fix the form of these functions we impose the conducting boundary
conditions on the disks. In particular, if the disks are held at fixed
potentials $\Delta_i$, then we will find a set of dual integral equations
similar in form to those in \cite{lmsvv,lsv,sneddon}. The conditions at the
$i^{th}$ disk are that for $r<R_i$
\begin{equation} \label{rsmR}
\int_0^\infty \frac{du}{u} J_0(ur)
	\left[A_i(u)+\sum_{j\neq i}A_j(u)e^{-u|d_j-d_i|} \right] =
	\Delta_i+2d_i^2-r^2 \, ,
\end{equation}
and for $r>R_i$
\begin{equation} \label{rlgR}
\int_0^\infty du J_0(ur) A_i(u) = 0 \, .
\end{equation}
We can make the ansatz
\begin{equation} \label{ansatz}
A_i(u)=\frac{2u}{\pi}\int_0^{R_i} dt \, \cos(ut) f_i(t) \, ,
\end{equation}
so that the conditions in \eqref{rlgR} are automatically satisfied, and the
conditions \eqref{rsmR} become
\begin{equation} \label{inteqsk}
f_i(r) + \sum_{j\neq i}\int_0^{R_j} dx \, \bar K_{ij}(x,r)f_j(x) = g_i(r) \, ,
\end{equation}
where
\begin{equation} \label{kbdef}
\bar K_{ij}(x,r) = \frac{|d_i-d_j|}{\pi}
	\left[
	\frac{1}{(x+r)^2+|d_i-d_j|^2}+\frac{1}{(x-r)^2+|d_i-d_j|^2}
	\right] ,
\end{equation}
$g_i(r)= \beta_i -2r^2$, and $\beta_i = \Delta_i+2d_i^2$.
Since the $g_i$ are all symmetric functions, it is simpler to take the system
as
\begin{equation} \label{inteqs}
f_i(r) + \sum_{j\neq i}\int_{-R_j}^{R_j} dx \, K_{ij}(x,r)f_j(x) = g_i(r) \, ,
\end{equation}
where
\begin{equation} \label{kdef}
K_{ij}(x,r) = \frac{1}{\pi}
        \frac{|d_i-d_j|}{(x-r)^2+|d_i-d_j|^2} \, .
\end{equation}
It is straightforward to show that the charge densities on the disks are given
in terms of the $f_i$ as
\begin{equation} \label{sigma}
\sigma_i(r) = \frac{W_0}{\pi^2} \left[ \frac{f_i(R_i)}{\sqrt{R_i^2-r^2}}
	-\int_r^{R_i} du\frac{f_i'(u)}{\sqrt{u^2-r^2}}\right] ,
\end{equation}
and that the total charges are
\begin{equation} \label{charge}
Q_i = \frac{W_0}{\pi} \int_{-R_i}^{R_i} du \, f_i(u) \, .
\end{equation}
To find the $f_i$ we must solve the set of linear equations in \eqref{inteqs},
schematically this takes the form
\begin{equation} \label{scheme}
\begin{pmatrix}
1 & K_{12} & \cdots \\
K_{21} & 1 & \cdots \\
\vdots & \vdots & \ddots
\end{pmatrix}
\begin{pmatrix}
f_1\\
f_2\\
\vdots
\end{pmatrix}=
\begin{pmatrix}
\beta_1-2r^2\\
\beta_2-2r^2\\
\vdots
\end{pmatrix} .
\end{equation}
Due to the complicated form of the kernels $K_{ij}$ this system is not easy
to solve analytically, however, it is straightforward to solve it numerically
using the Nystr\"om method (see, e.g.~\cite{numbook}). This consists of
discretizing the interval and solving the resulting linear system.
An additional set of constraints comes from ensuring that the charge densities
vanish at the edges of the disks. This amounts to enforcing that $f_i(R_i)=0$.
We define $f_i^{(j)}$ as the set of solutions to \eqref{inteqs} with
$g_i(r)=\delta_i^j$, where if there are $N$ disks $j=1,\ldots,N$, and 
$f_i^{(0)}$ as the set with $g_i(r)=2r^2$. The condition that the charge
density vanishes at the edge of the disk is then that
\begin{equation} \label{beta}
\sum_{j} f_i^{(j)}(R_i)\beta_j = f_i^{(0)}(R_i) \, .
\end{equation}
There will be a unique solution for $\beta_i$ if $\det(f_i^{(j)}(R_i))\neq 0$.
The full solution $f_i$ is then
\begin{equation} \label{fsoln}
f_i(r) = -f_i^{(0)}(r)+\sum_j\beta_j f_i^{(j)}(r) \, ,
\end{equation}
with the potentials $\phi_i$ given by
\begin{equation}
\phi_i = \int_{-R_i}^{R_i} dt \, G_i(r,z,t) f_i(t) \, ,
\end{equation}
where
\begin{equation} \label{gdef}
G_i(r,z,t) = \frac{1}{\pi}
	\frac{1}{\sqrt{(|z-d_i|+it)^2+r^2}} \, .
\end{equation}

We have therefore reduced the electrostatics problem to a very simple linear
system. In the case that there are only two disks, the problem is very
straightforward and solution has been used to understand the relationship
between SYM on $R\times S^2$ and little string theory \cite{lsv}.

\subsection{General solutions dual to the PWMM} \label{gpwmm}
Another theory for which Lin and Maldacena found the corresponding
electrostatics configurations is the plane wave matrix model \cite{bmn}. The
plane wave matrix model can be found by a consistent truncation of $\Ncal=4$
SYM on $R\times S^3$ to the set of constant modes on the sphere \cite{kkp}. The
vacua of matrix model are given by the scalars that come from the former
$\Ncal=4$ gauge field taking values in a representation of $SU(2)$ \cite{bmn}.
Lin and Maldacena \cite{lm} associated these vacua with configurations of
charged conducting disks above an infinite conducting plane.

The method of solution is very similar to the case above. For the sake of
brevity we will give the final solution. We will write the potential as
\begin{equation} \label{ppot}
V=V_0 \left( r^2 z -\frac23z^3 + \sum_i \phi_i(r,z) \right) ,
\end{equation}
where the first two terms are the background field and the $\phi_i$ arise from
the charged disks as
\begin{equation} \label{pphi}
\phi_i(r,z) = \int_{-R_i}^{R_i} dt \, G_i(r,z,t)f_i(t) \, .
\end{equation}
The Green function is
\begin{equation} \label{pgdef}
G_i(r,z,t) = \frac{1}{\pi} \Bigl(
	\frac{1}{\sqrt{(|z-d_i|+it)^2+r^2}}
	-\frac{1}{\sqrt{(|z+d_i|+it)^2+r^2}}
		\Bigr) ,
\end{equation}
and $f_i$ is a solution of the integral equation
\begin{equation} \label{pinteqs}
f_i(r) + \sum_{j}\int_{-R_j}^{R_j} dx \, K_{ij}(r,x) f_j(x) = g_i(r) \, ,
\end{equation}
in which the kernel is given by
\begin{equation} \label{pkdef}
K_{ij}(x,r) = \frac{1}{\pi} \left[
	\frac{|d_i-d_j|}{(x-r)^2+|d_i-d_j|^2} 
	-\frac{|d_i+d_j|}{(x-r)^2+|d_i+d_j|^2}
	\right] ,
\end{equation}
and $g_i(r)=\beta_i-2d_i r^2$, where $\beta_i=\Delta_i+\tfrac23 d_i^3$.
The differences between this solution and the one presented in section \ref{gen}
arise from the presence of the infinite conducting plane that, via the
method of images, implies the presence of oppositely charged conducting disks
below the image plane. However, the conditions on the charges on the disks
\eqref{sigma},\eqref{charge}, still hold, so the requirement that the charge
density vanishes at the edge of the disk is that $f_i(R_i)=0$. We may again
consider solutions to \eqref{pinteqs} in which $g_i(r)=\delta_i^j$, which we
will call $f_i^{(j)}$, and $f_i^{(0)}$ for which $g_i(r)=2d_i r^2$. The
condition that the charge density vanishes at the edge of each disk is again
\eqref{beta}, and $f_i$ will be given by \eqref{fsoln}.

As in the case of SYM on $R\times S^2$, the electrostatics problem has been
reduced to a very simple linear system. We will now show how we can solve this
system to study instantons at strong coupling on the field theory side
using this method.

\section{Instanton calculations} \label{inst}
Recently, Lin \cite{lin} has considered tunnelling between vacua in the plane
wave matrix model and in maximally supersymmetric Yang-Mills on $R\times S^2$.
It is possible to study this on both the gauge theory and gravity sides. In the
gauge theory case, this can be approached by directly studying the instanton
solutions \cite{yy}. Lin \cite{lin} has also shown that it is possible to
introduce a superpotential that gives a bound for the instanton action according
to
\begin{equation} \label{sandw}
S_{inst} = - \frac{1}{g^2} \Delta W \, .
\end{equation}
Lin \cite{lin} further studied this on the gravity side. Explicit answers were
given in the case that the disks in the corresponding electrostatics problem
were small and could be approximated by point charges. Moreover, Lin \cite{lin}
gave a prescription for how the instanton action could be expressed in terms of
the electrostatics potential in more general cases, but was not able to give
explicit expressions.

For completeness, we will briefly review Lin's prescription for finding the
instanton action from the gravity side \cite{lin}, and then we will demonstrate
the use of the techniques developed above to calculate the instanton action for
some non-trivial electrostatics configurations.

\subsection{Instantons on the Gravity side} \label{igen}
\begin{figure}
\includegraphics[scale=0.85]{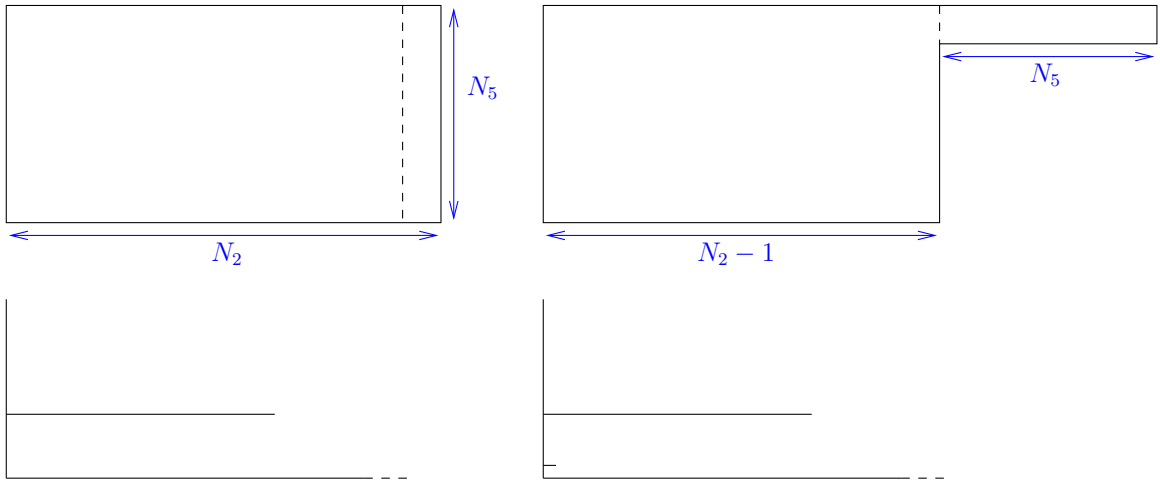}
\caption{The Young diagrams associated with the initial and final vacua in the
plane wave matrix model, and the electrostatics problems for the dual
supergravity solutions.}
\label{pwfig}
\end{figure}

Since Lin and Maldacena \cite{lm} have found electrostatics configurations
corresponding to $SU(2|4)$ symmetric field theory vacua, it is interesting to
understand how instantons in the field theories can be described in the gravity
picture. Lin \cite{lin} has studied this question by first considering vacua
in the field theories for which the electrostatics configurations do not differ
drastically. These instantons can be addressed by calculating the action for a
Euclidean D2-brane wrapping a non-contractible $\Sigma_3$ in the geometry. As
discussed in \cite{lm}, since the brane will be wrapping a cycle carrying some
$N_5$ units of flux, it should have $N_5$ D0-branes ending on it, and therefore
describe the creation of $N_5$ D0-branes in the throat, see figure \ref{pwfig}.

Using the mapping to an electrostatics configuration, the action for the
Euclidean D2-brane can be expressed in terms of the solution to the
electrostatics problem. Consider the case of a charged conducting disk at a
position $z_0$ above an infinite conducting plane, as shown in figure
\ref{pwfig}. Lin \cite{lin} has shown that the action for such a configuration
takes the form
\begin{equation} \label{se}
S_E = -\frac{2}{\pi}[V(z_0)-V(0)-z_0 V'(0)] \, ,
\end{equation}
where $V$ is the electrostatics potential evaluated along $r=0$, and prime
denotes differentiation with respect to $z$. This expression, however, is
proportional to the change in energy of the electrostatics configuration,
$S_E = 8\Delta U/\pi^3$. This led Lin \cite{lin} to identify the superpotential
at strong coupling as
\begin{equation} \label{wdef}
W \equiv -\frac{8g^2}{\pi^3} U = -\frac{16g^2}{\pi^3}\sum_i Q_i V_i \, ,
\end{equation}
where $U$ is the energy of the electrostatics configuration.

In the case that the disks are small relative to their separation, which is
at weak coupling in the gauge theory, the superpotential is given by the energy
of a system of point charges. Using the prescription \eqref{wdef}, Lin found
 \cite{lin}
\begin{equation} \label{linres}
W = \frac13 \sum_i N_2^{(i)} N_5^{(i) 3} \, ,
\end{equation}
in perfect agreement with the weak coupling gauge theory results \cite{lin,yy}.
In the case that the coupling is not weak, the charges arrange to form
extended disks. We will find the superpotential at strong coupling by solving
the electrostatics problem for a set of extended disks.

\subsection{Instantons in the PWMM} \label{PWMM}
In this section we will consider instantons in the simplest non trivial
electrostatics configuration, that of a single conducting disk carrying
$\pi^2 N_2/8$ units of charge at a distance $\pi N_5/2$ above a conducting
plane. This corresponds to a field theory vacuum with $N_2$ copies of the $N_5$
dimensional representation. We will determine the superpotential for
arbitrary $N_2$ and $N_5$, and calculate the action for Euclidean D2-brane
wrapping the non-contractible $\Sigma_3$.

In the case of small changes to the background mediated by the Euclidean
D2-brane, this will compute the action between a vacuum of the PWMM with $N_2$
copies of the $N_5$ dimensional representation, and a vacuum with $N_2-1$ copies
of the $N_5$ dimensional representation and $N_5$ copies of the trivial
representation. See figure \ref{pwfig}.

The electrostatics problem in this case can be approached using the technique
outlined in section \ref{gpwmm} applied to the case of a single disk above the
conducting plane. We consider a disk of radius $R$, at a distance $d$ from the
plane, which is a generalization of the approach in \cite{lmsvv}. The solution
to the electrostatics problem will be
\begin{equation} \label{pwv}
V=V_0 \left( r^2 z - \frac23 z^3  + \phi \right) ,
\end{equation}
where $\phi$ is
\begin{equation} \label{pwphi}
\phi(r,z) = \int_{-R}^{R} dt \, G(r,z,t) f(t) \, ,
\end{equation}
with Green function
\begin{equation} \label{pwG}
G(r,z,t) = \frac{1}{\pi} \left( \frac{1}{\sqrt{(|z-d|^2+it)^2+r^2)}}
	-\frac{1}{\sqrt{(|z+d|^2+it)^2+r^2)}} \right) .
\end{equation}
Here $f$ satisfies the integral equation
\begin{equation} \label{pwf}
f(r) + \int_{-R}^R dx \, K(r,x) f(x) = g(r) \, ,
\end{equation}
with
$g(r)=\beta-2dr^2$, $\beta=\Delta+\frac23 d^3$, and kernel
\begin{equation} \label{pwK}
K(r,x) = -\frac{1}{\pi} \frac{2d}{(x-r)^2+4d^2} .
\end{equation}
We will solve the problem by finding a numerical solution to the integral
equation \eqref{pwf}.

Solving this integral equation is straightforward. As a check on our numerical
results, we ensured that the asymptotic form for the superpotential in the limit
that the number of units of charge on the disks was small is given
by \eqref{linres}. Indeed, we found that for $N_5 \gg \lambda^\frac13 \gg 1$
\footnote{Here $\lambda\equiv g^2 N_2$, where $g$ is the Yang-Mills coupling of
the matrix model.}
\begin{equation} \label{asympsm}
W \approx \frac13 N_2 N_5^3 +
	a \sqrt\lambda N_2 N_5^{\frac32} \, ,
\end{equation}	
where the numerical constant $a\approx 1.4$.

\begin{figure}
\begin{center}
\begin{picture}(0,0)%
\includegraphics{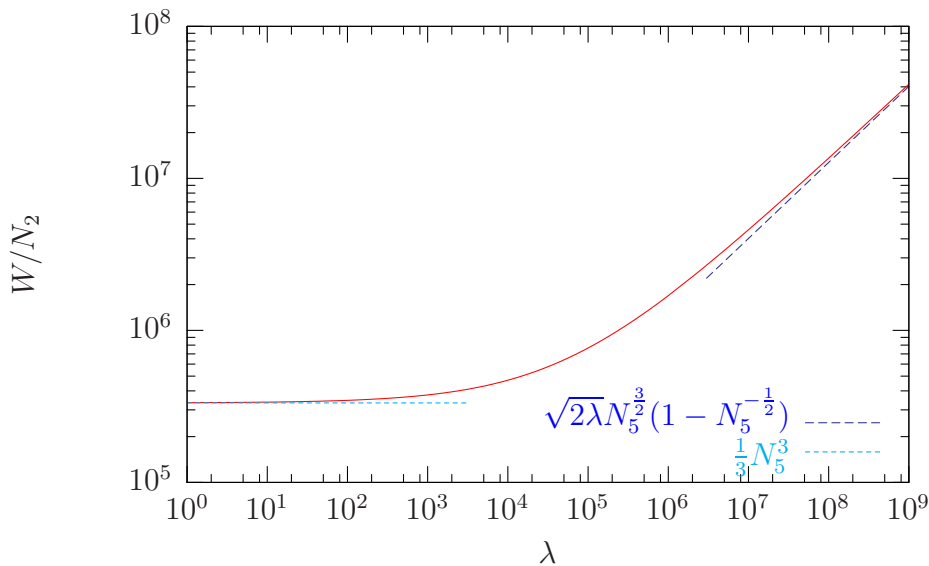}%
\end{picture}%
\begingroup
\setlength{\unitlength}{0.0200bp}%
\begin{picture}(18000,10800)(0,0)%
\put(3300,1650){\makebox(0,0)[r]{\strut{}$10^{5}$}}%
\put(3300,4517){\makebox(0,0)[r]{\strut{}$10^{6}$}}%
\put(3300,7383){\makebox(0,0)[r]{\strut{}$10^{7}$}}%
\put(3300,10250){\makebox(0,0)[r]{\strut{}$10^{8}$}}%
\put(3575,1100){\makebox(0,0){\strut{}$10^{0}$}}%
\put(5086,1100){\makebox(0,0){\strut{}$10^{1}$}}%
\put(6597,1100){\makebox(0,0){\strut{}$10^{2}$}}%
\put(8108,1100){\makebox(0,0){\strut{}$10^{3}$}}%
\put(9619,1100){\makebox(0,0){\strut{}$10^{4}$}}%
\put(11131,1100){\makebox(0,0){\strut{}$10^{5}$}}%
\put(12642,1100){\makebox(0,0){\strut{}$10^{6}$}}%
\put(14153,1100){\makebox(0,0){\strut{}$10^{7}$}}%
\put(15664,1100){\makebox(0,0){\strut{}$10^{8}$}}%
\put(17175,1100){\makebox(0,0){\strut{}$10^{9}$}}%
\put(550,5950){\rotatebox{90}{\makebox(0,0){\strut{}$W/N_2$}}}%
\put(10375,275){\makebox(0,0){\strut{}$\lambda$}}%
\put(14950,2975){\makebox(0,0)[r]{\strut{}\textcolor{blue}{$\sqrt{2\lambda}N_5^\frac32(1-N_5^{-\frac12})$}}}
\put(14950,2125){\makebox(0,0)[r]{\strut{}\textcolor{cyan}{$\frac13N_5^3$}}}
\end{picture}%
\endgroup
\end{center}
\caption{The superpotential for $N_5=100$. The dashed lines indicate
the asymptotic values for small and large $\lambda$ compared to $N_5$.}
\label{Wfig}
\end{figure}
The solution of the electrostatics problem when the disks are large gives
the superpotential at strong coupling. A plot of the superpotential for
$N_5=100$ is given in figure \ref{Wfig}. It is possible to extract the
asymptotic form for the superpotential in the limit that
$\lambda^\frac14\gg N_5 \gg 1$. We find the result
\begin{equation} \label{wasymp}
W \approx b \sqrt\lambda N_2 N_5^\frac32(1-N_5^{-\frac12})
	+ c \lambda^\frac14 N_2 N_5^2(1-N_5^{-\frac34}) \, ,
\end{equation}
where the numerical constants $b\approx1.4142\approx\sqrt2$, and $c\approx 0.7$.
In defining the superpotential, there was the freedom to choose an overall
constant factor. Here we have defined the superpotential to be zero for the
vacua given by $N_2 N_5$ copies of the trivial representation.

\begin{figure}
\begin{center}
\begin{picture}(0,0)%
\includegraphics{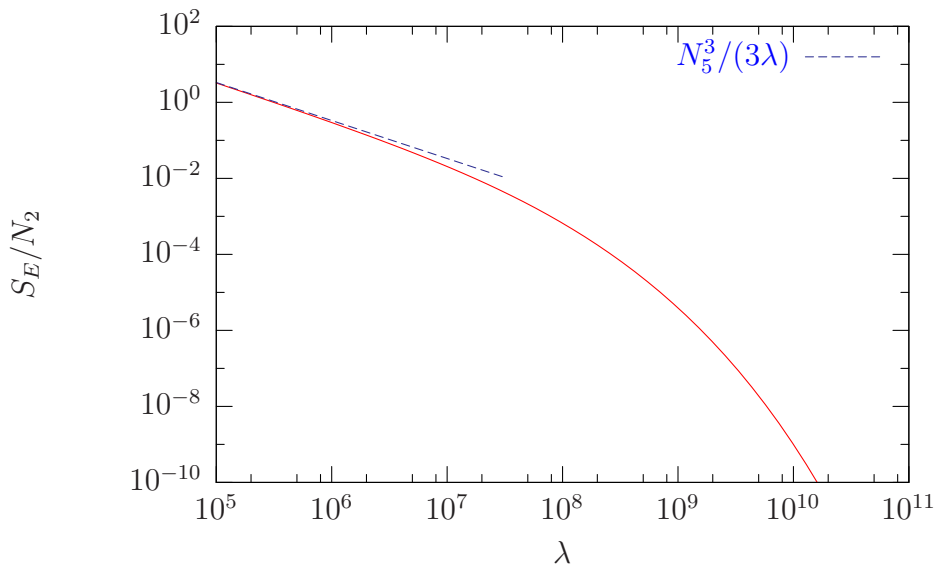}%
\end{picture}%
\begingroup
\setlength{\unitlength}{0.0200bp}%
\begin{picture}(18000,10800)(0,0)%
\put(3850,1650){\makebox(0,0)[r]{\strut{}$10^{-10}$}}%
\put(3850,3083){\makebox(0,0)[r]{\strut{}$10^{-8}$}}%
\put(3850,4517){\makebox(0,0)[r]{\strut{}$10^{-6}$}}%
\put(3850,5950){\makebox(0,0)[r]{\strut{}$10^{-4}$}}%
\put(3850,7383){\makebox(0,0)[r]{\strut{}$10^{-2}$}}%
\put(3850,8817){\makebox(0,0)[r]{\strut{}$10^{0}$}}%
\put(3850,10250){\makebox(0,0)[r]{\strut{}$10^{2}$}}%
\put(4125,1100){\makebox(0,0){\strut{}$10^{5}$}}%
\put(6300,1100){\makebox(0,0){\strut{}$10^{6}$}}%
\put(8475,1100){\makebox(0,0){\strut{}$10^{7}$}}%
\put(10650,1100){\makebox(0,0){\strut{}$10^{8}$}}%
\put(12825,1100){\makebox(0,0){\strut{}$10^{9}$}}%
\put(15000,1100){\makebox(0,0){\strut{}$10^{10}$}}%
\put(17175,1100){\makebox(0,0){\strut{}$10^{11}$}}%
\put(550,5950){\rotatebox{90}{\makebox(0,0){\strut{}$S_E/N_2$}}}%
\put(10650,275){\makebox(0,0){\strut{}$\lambda$}}%
\put(14950,9675){\makebox(0,0)[r]{\strut{}\textcolor{blue}{$N_5^3/(3\lambda)$}}}
\end{picture}%
\endgroup
\end{center}
\caption{The action for a euclidean D2-brane for the situation shown in
figure \ref{pwfig}. Here $N_5=100$. The dashed line shows the
asymptotic behaviour for $\lambda$ small compared to $N_5$.}
\label{sefig}
\end{figure}

We can also use the electrostatics solution to determine the action according
to \eqref{se} for the instanton shown in figure \ref{pwfig}. A plot of the
result for $N_5=100$ is shown in figure \ref{sefig}.
When $\lambda$ is small compared to $N_5$,
\begin{equation}
S_E \approx \frac13 \frac{N_2 N_5^3}{\lambda} \, ,
\end{equation}
and it falls off faster than any power of $\lambda$ when $\lambda$ is large.
When $\lambda$ is small compared to $N_5$, the potential on the disk in the
electrostatics problem is negative, which is due to the form of the background
potential. As the size of the disk increases the potential on the disk
increases. The instanton action according to \eqref{se} begins to fall off from
the behaviour at weak coupling near where the potential on the disk crosses
zero. It is sensible that it should vanish when the coupling is infinite, since
in that case we would expect the electric field to become constant between the
disks near the origin, and so the potential difference and dipole contributions
should cancel out. Let us briefly mention when these results should be valid.
The euclidean brane approximation will be valid when the number of units of
charge on the disk is large, the potential from the dipole at the origin is
small at the surface of the disk, and the curvature is small. Therefore
$\lambda$, $N_2$ and $N_5$ must all be large.

It would be helpful to better understand the scaling behaviour that was found
in \eqref{asympsm} and \eqref{wasymp}. We have taken the normalization of the
superpotential thus far to allow for direct comparison of vacua with $N_2$
copies of the $N_5$ dimensional representation to ones with $N_2N_5$ copies of
the trivial representation. The asymptotic behaviours we found in
\eqref{asympsm} and \eqref{wasymp} using this prescription have some interesting
features. In particular, at a fixed order in $\lambda$, we found that the
coefficients for the subleading terms in $N_5$ are one (i.e.\ the factors of
$(1-N_5^{-\alpha})$, where $\alpha$ is some positive number). The reason for
this is as follows. If we took the superpotential to be normalized to zero for
the empty background instead, it would have been advantageous to take out a
further scaling factor of $N_5^3\sim d^3$ in \eqref{pwv}. In that case, the
potential would be of the form $V=V_0 N_5^3 \bar V(R/d)$. Likewise, the charge
on the disk would then have the form $Q=V_0 N_5^4 \bar q(R/d)$. Since the total
charge is proportional to $N_2$, we must have that
\begin{equation} \label{qscale}
Q \sim N_2 \sim \frac{N_5^4}{g^2} \bar q(R/d) \, ,
\end{equation}
or
\begin{equation} \label{qcomb}
\bar q(R/d) \sim \frac{g^2 N_2}{N_5^4} = \frac{\lambda}{N_5^4} \, .
\end{equation}
We see, then, that functions of $R/d$ in the scaled electrostatics problem can
only depend on the combination of gauge theory parameters $\lambda/N_5^4$.
Therefore the superpotential with this alternative normalization must have the
form
\begin{equation} \label{wscale}
\bar W(\lambda,N_2,N_5) = N_2 N_5^3 \bar w\left( \frac{\lambda}{N_5^4} \right) .
\end{equation}
Our numerical results confirm this. We find the following asymptotic behaviour
for $\bar w$:
\begin{align} \label{fasymp}
\bar w(x) &\approx \frac 13\left(1 - \frac{10}{3} x^\frac23\right),
	& x &\ll 1\, , \\
\bar w(x) &\approx -\sqrt{2}x^\frac12 + 0.7 x^\frac14, & x &\gg 1 \, . \nonumber
\end{align}
The superpotential with the original normalization is given in terms of $\bar W$
by
\begin{equation} \label{normrelate}
W(\lambda,N_2,N_5)=\bar W(\lambda,N_2,N_5)-\bar W(\lambda N_5, N_2 N_5,1) \, .
\end{equation}
If we combine the contributions from the asymptotic behaviour of each of the
terms in this expression that come from the behaviour found in \eqref{fasymp},
then we will recover the asymptotic behaviour that was found in \eqref{asympsm}
and \eqref{wasymp}. The factors of $(1-N_5^{-\alpha})$ occur as a result of
those combinations.

\section{Discussion} \label{discussion}
In this paper we have given a prescription for finding the supergravity
solutions dual to general vacua of the plane wave matrix model and maximally
supersymmetric Yang-Mills on $R\times S^2$ by using the mapping of Lin and
Maldacena \cite{lm} on to axisymmetric electrostatics problems.

The prescription extends the technique developed in \cite{lmsvv,lsv} to
arbitrary electrostatics configurations. The electrostatics problems are
reduced to a set of integral equations that can be solved quite
straightforwardly using the Nystr\"om method.

We have shown that an application of the prescription to a specific case can
be used to study instantons at strong coupling in the plane wave matrix model.
In particular we found that the instanton action for a transition between
a vacuum described by $N_2$ copies of the $N_5$ dimensional representation
and one by $N_2-1$ copies of the $N_5$ dimensional representation and $N_5$
copies of the trivial representation falls off faster than any power of
$\lambda$ at strong coupling (see figure \ref{sefig}). We also
found that at strong coupling the superpotential for a vacuum with $N_2$ copies
of the $N_5$ dimensional representation behaves like
$\sqrt{2\lambda}N_2 N_5^{3/2}$, when $N_5$ is large (see figure \ref{Wfig}).
This demonstrates that the techniques developed above are useful for obtaining
strong coupling results in the field theory.

One question that would be interesting to address using this method is to
calculate the superpotential explicitly for more general electrostatics
configurations. For example, studying the vacuum of maximally supersymmetric
Yang-Mills theory on $R\times S^2$ in which
$\Phi=\diag(n,\ldots,n,-n,\ldots,-n)$, a similar scaling could be applied as
was done for the plane wave matrix model. In that case we would expect that
the corrections to the weak coupling results given by Lin \cite{lin} would
depend on the parameter $\lambda/n^3$. It would certainly be interesting to
study that case in detail, as well as other more general vacua.

One open question is to prove that requiring the charge density vanish at the
edge of each disk implies that there is a unique solution to the electrostatics
problem. The condition for a unique solution to exist is given above by
requiring $\det(f_i^{(j)}(R_i))\neq 0$. We have not been able to prove that
this is true in general. It would be interesting to do so.

Finally, it would be very interesting to use this method for finding the
dual geometry to study other strong coupling phenomena on the gauge theory side.

\section*{Acknowledgments}
We would like to thank  Brian Shieh, Henry Ling and especially Mark Van
Raamsdonk for helpful discussions. This work has been supported in part by the
Natural Sciences and Engineering Council of Canada.

\end{document}